\documentclass[conference]{IEEEtran}
\pdfoutput=1
\IEEEoverridecommandlockouts 
\usepackage[backend=biber,style=ieee,dashed=false]{biblatex}
\usepackage{amsmath,amssymb,amsfonts,amsthm}

\newtheorem{theorem}{Theorem}
\newtheorem{assumption}{Assumption}
\newtheorem{property}{Property}
\usepackage{graphicx}
\usepackage{textcomp}
\usepackage{xcolor}
\usepackage{algorithm,algpseudocode}
\algrenewcommand\algorithmicrequire{\textbf{Input:}}
\def\BibTeX{{\rm B\kern-.05em{\sc i\kern-.025em b}\kern-.08em
		T\kern-.1667em\lower.7ex\hbox{E}\kern-.125emX}}
\usepackage{hyperref}

\usepackage{todonotes}
\setuptodonotes{fancyline, inline, color=blue!30}

\usepackage{multirow}

\usepackage[style=alttree]{glossaries}
\makeglossaries
\newglossaryentry{slot}
{
    name=slot,
    description={The unit of the voting power. During an epoch, blocks are produced by a set of elected validators consisting of $n$ slots. A single validator may occupy multiple slots.}
}

\newglossaryentry{epoch}
{
    name=epoch,
    description={An epoch is a segment of the blockchain produced by a fixed set of elected validators. It consists of multiple batches, each of which is finalized with a macro block.}
}

\newglossaryentry{batch}
{
    name=batch,
    plural=batches,
    description={A batch is a segment of an epoch, which consists of a fixed number of micro blocks and ends with a macro block.}
}

\newglossaryentry{macro-block}
{
    name={macro block},
    description={Macro blocks are produced with the Tendermint protocol at the end of each batch. The current slot owner functions as a proposer. Macro blocks do not contain transactions and are final.}
}

\newglossaryentry{micro-block}
{
    name={micro block},
    description={Micro blocks make up the largest part of the blockchain. They are signed by the selected slot owner and contain transactions.}
}

\newglossaryentry{skip-block}
{
    name={skip block},
    description={A skip block is a special form of micro block that contains signatures from two-thirds of the slots. It is produced instead of a micro block if the original slot owner fails to produce the block.}
}

\newglossaryentry{slot-owner}
{
    name={slot owner},
    description={Slot owner is a synonym for the block producer/proposer. The slot owner is the slot selected from the current set of elected validators to produce a given block.}
}

\newglossaryentry{seed}
{
    name={random seed},
    description={The random seed is a source of randomness included in every block. It is produced by the slot owner from a verifiable random function.}
}

\newglossaryentry{elected-validator}
{
    name={elected validator},
    description={A member of the list of slots selected in an election block.}
}
\glsfindwidesttoplevelname

\newcommand{\ceil}[1]{\lceil {#1} \rceil}

\begin{document}

\title{Albatross\\
	\Large An Optimistic Consensus Algorithm\\
	\hspace{0pt} \\
	\normalsize April 16, 2024
	\thanks{This work is licensed under the Creative Commons Attribution-ShareAlike 4.0 International License.}
}

\author{
	\IEEEauthorblockN{Pascal Berrang}
	\IEEEauthorblockA{\textit{Nimiq Foundation}\\
	pascal@nimiq.com
	}
        \and
        \IEEEauthorblockN{Inês Cruz}
	\IEEEauthorblockA{\textit{Nimiq Foundation}\\
	ines.cruz@nimiq.com
	}
        \and
	\IEEEauthorblockN{Bruno Fran\c{c}a}
	\IEEEauthorblockA{\textit{Nimiq Foundation}\\
	bruno@franca.xyz
	}
	\and
	\IEEEauthorblockN{Philipp von Styp-Rekowsky}
	\IEEEauthorblockA{\textit{Nimiq Foundation}\\
	philipp@nimiq.com
	}
	\and
	\IEEEauthorblockN{Marvin Wissfeld}
	\IEEEauthorblockA{\textit{Nimiq Foundation}\\
	marvin@nimiq.com
	}
}

\maketitle
\thispagestyle{plain} 
\pagestyle{plain} 

\begin{abstract}
The consensus protocol is a critical component of distributed ledgers and blockchains. Achieving consensus over a decentralized network poses challenges to transaction finality and performance. Currently, the highest-performing consensus algorithms are speculative BFT algorithms, which, however, compromise on the transaction finality guarantees offered by their non-speculative counterparts.

In this paper, we introduce Albatross, a Proof-of-Stake (PoS) blockchain consensus algorithm that aims to combine the best of both worlds. At its heart, Albatross is a speculative BFT algorithm designed to provide strong probabilistic finality and high performance. We complement this by periodically establishing finality through the Tendermint protocol.
We prove our protocol to be secure under standard BFT assumptions and analyze its performance both on a theoretical and practical level. For that, we provide an open-source Rust implementation of Albatross. Our real-world measurements support that our protocol has performance characteristics close to the theoretical maximum for single-chain Proof-of-Stake consensus algorithms.
\end{abstract}

\section{Introduction}
\label{sec:intro}
The area of distributed ledgers is a vast and quickly developing landscape. At the heart of most distributed ledgers is their consensus protocol. The consensus protocol describes the way participants in a distributed network interact with each other to obtain and agree on a shared state. While classical Byzantine Fault Tolerant (BFT) algorithms are designed to work in closed, size-limited networks only, modern distributed ledgers -- and blockchains in particular -- often focus on open and permissionless networks.

The most famous classical consensus algorithm is PBFT, or Practical Byzantine Fault Tolerance~\cite{castro1999practical}. PBFT has greatly influenced the field since its creation in 1999 and has inspired many of the consensus algorithms being developed for blockchains. For example, one of the most widely used blockchain consensus algorithms is Tendermint~\cite{buchman2016tendermint,buchman2018latest}. Tendermint ensures state replication and provides a high degree of finality. However, consensus theory has evolved significantly, and nowadays, the BFT algorithms providing the highest performance are \emph{speculative} algorithms.

Speculative BFT refers to a class of algorithms that have two modes for consensus: (1) the \emph{optimistic} mode, when nodes are well-behaved, prioritizing speed, and (2) the \emph{pessimistic} mode, where the goal is to make progress despite of malicious nodes. Compared to traditional BFT algorithms, speculative algorithms are significantly faster due to their optimistic mode. This model enables a performance comparable to centralized systems. When misbehavior is detected, speculative BFT algorithms switch to the pessimistic mode and prevent invalid updates. Then, the protocol can resume using the optimistic mode again. During optimistic mode, speculative consensus algorithms rarely provide finality guarantees.

In this paper, we present Albatross, a Proof-of-Stake (PoS) blockchain consensus algorithm that aims to combine the best of both worlds. Albatross is designed to be a high-performing, speculative BFT algorithm for permissionless networks. During its optimistic execution, it aims to achieve strong probabilistic transaction finality. Moreover, it provides a periodic high degree of finality using Tendermint~\cite{buchman2016tendermint,buchman2018latest}. Albatross' unique design also enables synchronization with low hardware requirements.

In particular, our contributions are as follows:
\begin{enumerate}
    \item We propose a novel speculative consensus algorithm designed to provide strong probabilistic finality and periodic assurances of finality at regular intervals.
    \item We argue about the security of our algorithm and relate our security assumptions to the maximum amount of malicious stake we can support.
    \item We analyze the security implications of protocol upgrades in Albatross and explain how to generalize the results to other PoS blockchains.
    \item We briefly discuss how the design of our algorithm supports efficient syncing in constrained environments (e.g., browsers and smartphones).
    \item We provide an open-source Rust implementation of our algorithm.
    \item We present real-world performance measurements taken from the Nimiq PoS public testnet.
\end{enumerate}

\section{Protocol Overview}
\label{sec:overview}
In this section, we will give a summary of Albatross, describing the different types of validators, blocks, and the chain selection algorithm. We also present the protocol's behavior in the optimistic mode, discuss how validators may misbehave, and how such misbehavior is countered in our protocol.

We will focus on the general structure of Albatross and leave the details for Section~\ref{sec:specification}.

\subsection{Validators \& Stakers}
In Albatross, we distinguish between validators and stakers. 

A validator signals their willingness to participate in the protocol for block production and transaction validation by allocating a portion of their funds as stake.

Any validator that is currently elected to participate in the block production becomes an \emph{\gls{elected-validator}}. The higher its total stake, the higher the probability of being elected.

A participant who does not have the resources to become a validator can instead sign up as a staker. A staker delegates its funds to a validator, which produces and validates blocks on its behalf.

\subsection{Blockchain Structure}
\begin{figure}[t]
    \centering
    \includegraphics[width=\linewidth]{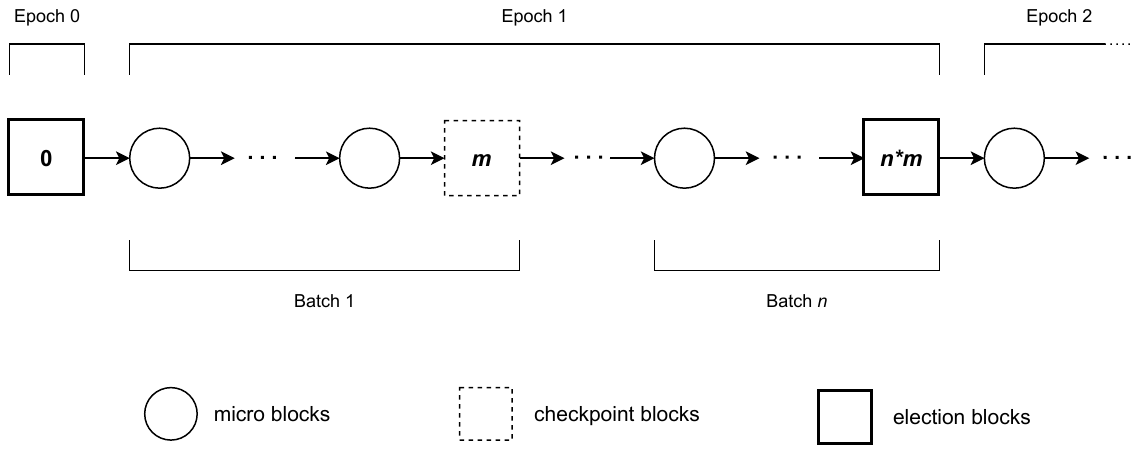}
    \caption{Blockchain structure}
    \label{fig:blocks}
\end{figure}

As illustrated in \autoref{fig:blocks}, we divide the blockchain into \glspl{epoch}, which are sections of blocks produced by a fixed set of validators.
Each epoch is split into $n$ \glspl{batch} of $m$ blocks. A batch provides finality for the respective segment.

To achieve finality and changes in the set of validators, we introduce a type of block called \emph{\gls{macro-block}} that is produced as the last block of each batch. The remaining blocks are called \emph{\glspl{micro-block}} and contain user-generated transactions.

\begin{itemize}
    \item \textbf{Macro blocks:} 
    Macro blocks are produced with Tendermint, where a random validator is chosen to propose the new macro block.
 
    Macro blocks at the end of an epoch elect a new set of validators and are called \emph{election blocks}. The remaining macro blocks only provide finality and are called \emph{checkpoint blocks}.
    \item \textbf{Micro blocks:} These blocks are the ones that contain user-generated transactions. Each micro block is produced by a randomly chosen \gls{elected-validator} and contains the transactions to be included. Micro blocks only need to be signed by the corresponding block producer.
\end{itemize}

Additionally, all blocks also contain a \gls{seed} that is used to select the next block producer.

\subsection{Speculative Block Production}
When all validators behave honestly, the protocol can run in \emph{optimistic mode}. This means that the chain progresses without any forks. 
Nevertheless, Albatross also needs to be able to withstand malicious validators.

\subsubsection{Optimistic Mode}
In this mode, every block is produced by a single active validator. 
We deterministically select the block producer from the list of \glspl{elected-validator} using the \gls{seed} of the predecessor block. This block producer is called the \emph{\gls{slot-owner}}.

\subsubsection{Pessimistic Mode}
When malicious behavior is identified, the protocol enters pessimistic mode. This mode is designed to handle forks, block delays, and invalid blocks.

\paragraph{Forks}
Due to the properties of Tendermint, no two \glspl{macro-block} can be finalized at the same block height.
In contrast, for \glspl{micro-block}, forks can be created if a validator produces more than one block at the same block height.
To deal with this, we propose \emph{fork proofs}. Anyone can create a fork proof. Two block headers at the same block height, signed by the same slot owner, constitute sufficient proof.

If a malicious validator creates or continues a fork, an honest validator will eventually produce a block containing a fork proof, and the malicious validator will be punished.
\paragraph{Double Votes and Double Proposals}
Despite Tendermint's strong probabilistic finality, a malicious validator can still interfere by double voting or double proposing. To address this issue, we propose the introduction of \emph{double vote proofs} and \emph{double proposal proofs}. Similarly to fork proofs, any honest validator can produce a block containing these proofs. As their structure is specific to Tendermint's protocol implementation, we define the structure of these proofs adapted to our implementation in ~\autoref{sec:equivocations}.

\paragraph{Block Delays}
When a validator does not produce a micro block in the expected time, Albatross allows the remaining validators to produce a \emph{\gls{skip-block}} instead. It advances the blockchain by one block and skips the misbehaving validator as the block producer.

A skip block is a micro block with some differences: it does not hold transactions, and its random seed is copied from the predecessor block. It requires two-thirds of the \glspl{elected-validator} to sign it in order to be accepted.
Once included in the chain, the delaying validator will be punished.

This process is similar to the view-change protocol in PBFT~\cite{castro1999practical}.

\paragraph{Invalid blocks}
When a validator produces an invalid block, the other validators will ignore that block and prevent it from propagating through the network. The validators may additionally ban that peer on the network level.

This behavior is more efficient than attempting to punish directly since verifying the proof of misbehavior is more costly than producing invalid blocks. Moreover, it may not be possible to punish the misbehaving party if the signature does not match any validator. 
Still, producing an invalid block can lead to an indirect punishment through a block delay.

\paragraph{Punishment}
We propose to distinguish severe protocol violations, namely fork producers, double voters, and double proposers, from delayed block producers. Validators committing severe violations will be jailed, while those causing delays will face penalties.
Jailing consists of a defined period, measured in \glspl{epoch}, during which offending validators are banned from block production, unable to withdraw their funds and have their pending rewards burned. On the other hand, delaying validators are penalized by being deactivated and having their slot rewards burned. These validators remain deactivated until they signal their interest in rationally participating in the consensus.
It is possible to aggravate the punishments by slashing the validators’ stake. We leave this up to the protocol's concrete implementation.

\section{Preliminaries}
\label{sec:prelims}

\subsection{Tendermint}
\label{sec:prelims:tendermint}
\begin{figure}[t]
    \centering
    \includegraphics[width=\linewidth]{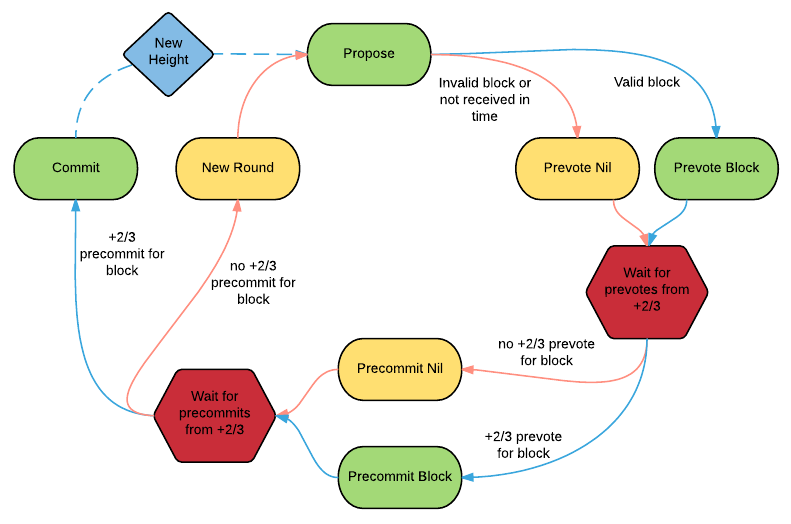}
    \caption{Tendermint phases (sourced from~\cite{buchman2016tendermint})}
    \label{fig:tendermint}
\end{figure}

Tendermint~\cite{buchman2016tendermint,buchman2018latest} is a Byzantine Fault Tolerant consensus algorithm based on a peer-to-peer gossip protocol. It is tailored towards a blockchain scenario and assumes the presence of validators who are proposing and voting on blocks.

The Tendermint protocol works in rounds, and each round has only one valid block proposer. To cope with network asynchrony or faulty proposers, a round can be skipped as long as less than one-third of the validators are offline or partitioned. Similarly to PBFT, after a block has been proposed, the validators go through two phases of voting before committing to a proposed block.

\autoref{fig:tendermint} depicts the different phases of the Tendermint protocol for a given block height. After the correct proposer has proposed the block, validators vote on the proposal in two phases: prevote and precommit. In each phase, validators must either vote for the block or against it (voting for \emph{nil}). The protocol can only progress after receiving votes from two-thirds or more (+2/3) of the validators. After having received two-thirds or more votes precommitting for the block, the block is said to be committed. If a proposal is not received in time, validators will vote \emph{nil} to move into the next round, resulting in a new proposer.

A timeout mechanism prevents indefinite delays in the decision-making for a block height within the protocol. Leaders have a timeframe to broadcast proposals, and if they fail to broadcast within this window, validators indicate not witnessing the proposal by voting nil. Consequently, the protocol transitions to a new round. Likewise, timeouts start during the prevote and precommit steps when the protocol accumulates more than two-thirds of votes in total (block and \emph{nil}) but fails to reach a definitive consensus for either. This allows for potential voting improvement while preventing the process from stalling while waiting for sufficient votes. As subsequent rounds progress, timeouts gradually increase, enhancing the probability of achieving consensus over an extended period. This adaptive adjustment minimizes the risk of consensus failure due to time constraints, thereby facilitating the protocol progress.

An additional locking mechanism prevents malicious coalitions of less than one-third from compromising the protocol's safety. This mechanism restricts the validators' votes depending on their previous prevotes and precommits at the same height, thus ensuring that no two blocks can be committed at the same height.

\subsection{Verifiable Random Functions}
\label{sec:prelims:vrf}
A verifiable random function (VRF~\cite{micali1999verifiable}) is a pseudo-random function (PRF~\cite{katz2020introduction}) that can provide a publicly verifiable proof that its output is correct.

More formally, after a user creates a public key $Y$ and a secret key $x$, given an input $s$, the user can calculate the VRF $\pi,r=\textsf{VRF}_x(s)$. Here, $r$ is the pseudo-random output, and $\pi$ is proof for the correct computation of the VRF. Then, anyone who knows the public key and the proof can verify that $r$ was correctly computed without learning the secret key $x$.

Verifiable random functions have three important properties:
\begin{itemize}
	\item \textbf{Pseudo-randomness:} The only way of predicting the output, better than guessing randomly, is to calculate the function.
	\item \textbf{Uniqueness:} For each input $s$ and secret key $x$ there is only one possible output $r$.
	\item \textbf{Public verifiability:} Anyone can verify that the output was correctly computed.
\end{itemize}

Possible instantiations of VRFs are BLS signatures~\cite{boneh2004short} or Signal's VXEdDSA signature~\cite{perrin2016xeddsa}.

\section{Protocol Specification}
\label{sec:specification}
We will now give the technical specification of Albatross, describing in detail validators and stakers, the block format, the misbehavior handling protocols, the rewards and punishments policies, and the chain selection algorithm.

\subsection{Validators \& Stakers}
Validators and stakers register on-chain with their respective stake. The registry also stores additional information, such as the stakers' delegation, the validators' public keys, and what state validators are in. 

A validator can be marked as active or inactive. Only active validators can be elected for block production. The protocol offers signaling transactions to the on-chain registry to switch between those states, as well as to create and delete validators and stakers.

\subsection{Validator Selection}
There are two cases in which we need to select a random subset of validators: at the end of an \gls{epoch} to elect new validators and before every \gls{micro-block} to select the next block producer.
We rely on a random beacon based on verifiable random functions to facilitate a pseudo-random but deterministic selection.

\subsubsection{Random Beacon}
In the \textit{genesis} block, there will be an initial \gls{seed}. This initial seed will need to be sourced from the outside world. We can use, for example, lottery numbers~\cite{baigneres2015million} or the hashes of Bitcoin headers~\cite{bonneau2015random}. Another possibility, to further reduce any possibility of bias in the initial seed, is to employ distributed randomness generation algorithms~\cite{syta2016randhound}.

In subsequent blocks, a new random seed is produced from the last seed using a VRF based on the block producer's secret key. This creates an infinite chain of random seeds that can be publicly verified. This process shares similarities with the use of VRFs in Algorand~\cite{chen2016algorand}.

\subsubsection{Validator Election}
In Albatross, a new list of $n$ \glspl{elected-validator} is chosen at the end of every epoch.
We refer to the positions of this list as \emph{\glspl{slot}}. 
The probability of being chosen for a slot is proportional to the total stake of a validator, which consists of its own and all its delegated stake. A validator can be selected for more than one slot.

We start by getting the addresses and the corresponding total stake of every active validator. An active validator is one who is eligible to receive slots for block production and is marked as active. Then, we order the addresses deterministically (for example, by lexicographic order). Lastly, we map the ordered addresses to their total stake, such that the amount represents a range. For example, if there are 10 tokens staked by $\textit{A}_1$, 50 tokens by $\textit{A}_2$, and 15 tokens by $\textit{A}_3$, then the mapping would be the one shown in \autoref{table:stake}.

\begin{table}[h]
	\caption{Validator stakes ordered and mapped to ranges}
	\label{table:stake}
	\centering
	\begin{tabular}{ c | c | c }
		Address & Stake & Range \\
		\hline
		$\textit{A}_1$ & 10 & $[0, 9]$ \\
		$\textit{A}_2$ & 50 & $[10, 59]$ \\
		$\textit{A}_3$ & 15 & $[60, 74]$ \\
	\end{tabular}
\end{table}

This mapping will then feed into \autoref{algorithm:validator-selection} to choose the new list of elected validators.
The algorithm makes use of a pseudo-random function (PRF~\cite{katz2020introduction}) $\mathit{prf}_K(x)$ with key $K$ and input $x$. We instantiate the key with the random seed $S$ obtained from the election block.

\begin{algorithm}
	\caption{Validator Election}
	\label{algorithm:validator-selection}
	\begin{algorithmic}[1]
		\State $\mathcal{V}_\textsf{elected} \gets [~]$
		\State $S \gets \text{random seed of election block}$
		\State $t \gets \text{total amount staked}$
            \State $n \gets \text{number of slots}$
		\State $i \gets 0$
		\While{$|\mathcal{V}_\textsf{elected}| < n$}
			\State $r \gets \mathsf{prf}_S(i) \mod t$
			\State $v \gets$ active validator whose range contains $r$
			\State $\Call{insert}{\mathcal{V}_\textsf{elected}, v}$
			\State $i \gets i + 1$
		\EndWhile
		\State \Return $\mathcal{V}_\textsf{elected}$
	\end{algorithmic}
\end{algorithm}

\subsubsection{Block Producer Selection}
Both micro and macro blocks require a block producer or proposer. This \gls{slot-owner} is determined by the random seed $S$ of the predecessor block, and the list of elected validators. The probability of being chosen as the slot owner is proportional to the number of slots of a given validator.

To efficiently calculate the slot owner, \autoref{algorithm:slot-owner} shuffles the list of well-behaving elected validators using the Fisher-Yates algorithm~\cite{fisher1963statistical,durstenfeld1964algorithm} and a pseudo-random number generator seeded with $S$. We then choose the slot owner based on an index. While for macro blocks, the index corresponds to the round number, for micro blocks, we fix the index to the block height. The index is required since Tendermint may require multiple rounds with different leaders and skip blocks carry over the previous seed. Using randomly selected validators at every block reduces the chances of predicting the next slot owner and thus prevents an attacker from taking over the chain.

\begin{algorithm}
	\caption{Slot Owner Selection}
	\label{algorithm:slot-owner}
	\begin{algorithmic}[1]
		\State $\mathcal{V}_\textsf{elected} \gets$ slots in current epoch
            \State $\mathcal{V}_\textsf{punished} \gets$ punished slots in current epoch
		\State $S \gets$ \gls{seed}
            \State $n \gets \text{number of slots}$
            \State $\mathit{index} \gets$ micro block or Tendermint round number
            \State
            \State $\mathcal{V} \gets \mathcal{V}_\textsf{elected}\setminus \mathcal{V}_\textsf{punished}$
		\State $\mathcal{V}_\textsf{shuffled} \gets \Call{FisherYatesShuffle}{\mathcal{V}, \mathsf{prng}(S)}$
		\State \Return $\mathcal{V}_\textsf{shuffled}[\mathit{index} \mod n]$
	\end{algorithmic}
\end{algorithm}

\subsection{Block Format}
\label{sec:block-format}
For the Albatross protocol, the minimal information needed to be stored in \glspl{micro-block} and \glspl{macro-block} is displayed in \autoref{table:blocks}.

\begin{table}[h]
	\caption{Block format}
	\label{table:blocks}
	\centering
	\begin{tabular}{|l|}
            \hline
		\multicolumn{1}{|c|}{\emph{\textbf{Micro Block}}} \\
		\hline
		\multicolumn{1}{|c|}{\textbf{Header}} \\
		\hline
            \emph{block number} \\
            \emph{random seed} \\
            \emph{body hash} \\
            \emph{parent hash} \\
            \\
            \\
		\hline
            \multicolumn{1}{|c|}{\textbf{Body}} \\
		\hline
            \emph{transactions}$^1$ \\
            \emph{equivocation proofs}$^1$ \\
            \hline
            \multicolumn{1}{|c|}{\textbf{Justification}} \\
		\hline
            \emph{signature of slot owner}\\ \textbf{or} \emph{skip block proof}\\
            \hline
            \multicolumn{1}{l}{}\\
            \multicolumn{1}{l}{$^1$empty in skip blocks}
	\end{tabular}
        \quad
        \begin{tabular}{| l |}
            \hline
		\multicolumn{1}{|c|}{\emph{\textbf{Macro Block}}} \\
		\hline
		\multicolumn{1}{|c|}{\textbf{Header}} \\
		\hline
            \emph{block number} \\
            \emph{random seed} \\
            \emph{body hash} \\
            \emph{parent hash} \\
            \emph{Tendermint round number} \\
            \emph{parent election hash} \\
		\hline
            \multicolumn{1}{|c|}{\textbf{Body}} \\
		\hline
            \emph{validators}$^2$ \\
            \\
            \hline
            \multicolumn{1}{|c|}{\textbf{Justification}} \\
		\hline
            \emph{Tendermint proof}\\
            \\
            \hline
            \multicolumn{1}{l}{}\\
            \multicolumn{1}{l}{$^2$empty in checkpoint blocks}
	\end{tabular}
\end{table}

The headers of micro and macro blocks only differ in that macro blocks contain the Tendermint round number and link to the parent election block. This link will allow us to simplify syncing the chain.
Since micro and macro blocks fulfill different purposes, their bodies and justifications do not share any fields.

Micro blocks contain transactions and fork proofs in their body, and their justification consists only of the identifier and signature of the \gls{slot-owner}.
There is a particular type of micro block, the \gls{skip-block}. A skip block has an empty body, and its justification aggregates the signatures of two-thirds of the \glspl{elected-validator}.

Macro blocks are produced with the Tendermint protocol (see~\autoref{sec:prelims:tendermint}). Validators either agree on the block proposal or start a new round with a new proposer until two-thirds accept the proposal. The justification contains the aggregate signature of the precommit step. Finally, election blocks include the list of elected validators for the next \gls{epoch}, while the body is empty for checkpoint blocks.

\subsection{Pessimistic Mode}
The ways Albatross handles validator misbehavior differ between \glspl{macro-block} and \glspl{micro-block}.
While macro blocks are secured through the Tendermint protocol, micro blocks need unique mechanisms for dealing with forks and block delays.

\subsubsection{Equivocation Proofs}
\label{sec:equivocations}
Equivocation refers to the act of a validator producing conflicting blocks, proposals, or double votes at the same round and step. Equivocation proofs detect malicious behavior and provide evidence for penalizing validators engaging in such behavior.

Validators have the option to submit proofs attesting to malicious behavior related to forks, double proposals, and double votes. A sufficiently large reporting window must be defined to report misbehavior. However, the potential full node's history pruning discussed in \autoref{sec:full-nodes} may limit the reporting window. We propose a reporting window of approximately two epochs to provide sufficient time for validators to identify, address, and submit proofs of malicious behavior to be included in the blockchain.

\paragraph{Fork Proofs}
To punish a \gls{slot-owner} that creates two or more micro blocks at the same block height, other validators can include a fork proof into the blockchain. This happens regardless of block validity. It is smaller and simpler to prove that any two blocks exist at the same height rather than proving the same for \emph{valid} blocks only.

The fork proof consists of two distinct micro block headers, their respective justifications, and the previous \gls{seed}. \Glspl{skip-block} are explicitly excluded from fork proofs. In order for the fork proof to be valid, the following conditions must be met: (1) the block headers must have the same block number and random seed; (2) the justifications must be valid; (3) the random seeds are correctly derived from the previous one.

\begin{figure}[t]
    \centering
    \includegraphics[width=\linewidth]{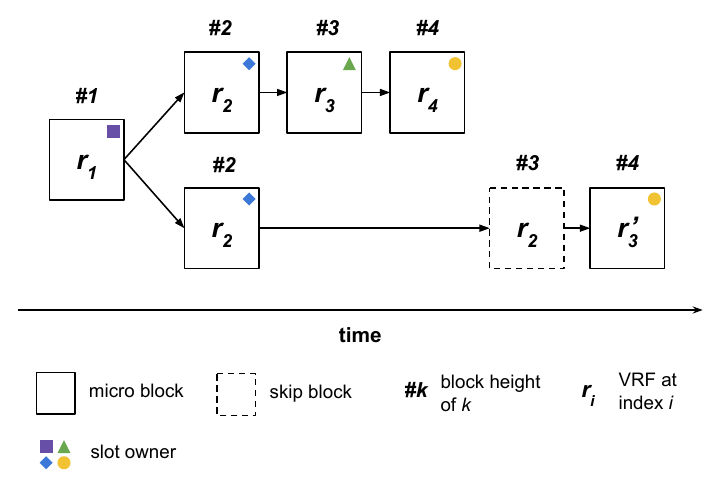}
    \caption{Legitimate fork at block height \#4}
    \label{fig:fork}
\end{figure}

The fork proof must include the random seed to verify the slot owner, even if the previous block is unknown.
Given that a fork starts from the same random seed $r_1$ and the VRF output is unique, the chain of random seeds for both subchains is usually the same.
However, if a skip block is added to one of the subchains, the seeds will begin to differ, as shown in \autoref{fig:fork}.
This may lead to a scenario in which a validator legitimately extends one of the subchains (block~\#4, upper chain) and is afterwards forced by a skip block on the other subchain to produce another block at the same height (block~\#4, lower chain).
While the skip block takes precedence and will resolve the fork, the validator involuntarily continued the fork and should thus not be punished.
That is why we require the random seeds in both headers of the fork proof to be the same.

\paragraph{Double Vote Proofs}
For each Tendermint proposal, there are two voting steps in which validators cast votes for block or \emph{nil}. When a validator casts votes for different proposals at the same round and step, it qualifies as a double vote, undermining the principles of Tendermint's voting mechanism—one vote per round and step. To address this misbehavior, double vote proofs are produced to detect and penalize validators engaging in such behavior.

The double vote proof structure includes essential verification components. These components contain the validator's address, the Tendermint identifier, and the two proposal hashes associated with the conflicting votes. The aggregated signatures are also included, but verifying them requires knowledge of the participating validators. A bitset of signers is included to facilitate this verification, enabling the identification of validators involved in signing the specific proposals. This serves as sufficient proof to pinpoint the malicious validator.

\paragraph{Double Proposal Proofs}
Similar to forks, double proposals can disrupt the integrity of the consensus process. In Albatross, macro blocks are produced with Tendermint. During each round, the designated leader is responsible for proposing a macro block, which is then broadcasted across the network. However, malicious validators may seek to tamper with the proposal by proposing two distinct macro headers at the same block height. To address this, honest validators may include double proposal proof to detect the malicious validators and thus penalize such behavior.

Validating double proposal proofs involves verifying the authenticity of the conflicting proposals and associated signatures. This proof includes the validator's address, the two macro headers from the same round proposed by the malicious validator, and the corresponding signatures for each header, providing evidence of the validator's actions.

\subsubsection{Skip Blocks}
\label{sec:specification:skip-blocks}
In optimistic mode, a single validator is enough to delay the block production. When a slot owner fails to produce its block on time, the other validators can vote on including a skip block instead.

Skip blocks are canonical because they reuse the random seed of the previous block and are otherwise empty. To be valid, a skip block needs to be signed by more than $\ceil{\frac{2}{3}n}$ \glspl{elected-validator}.

With each block added to the blockchain, validators start a timer for the production of the next block.
If they do not receive the next block within a predefined time window, validators will broadcast a signature for a skip block instead.
After receiving enough signatures, the skip block is immediately added to the chain.

Validators will continue aggregating signatures, even if they receive a block from the slot owner after the timeout. Once the threshold of $\ceil{\frac{2}{3}n}$ signatures is met, the skip block takes precedence over any block produced by the slot owner at the same block height.

\subsection{Rewards}
In Albatross, block production is extremely cheap; any regular computer with a good internet connection suffices. Consequently, validators do not need a large incentive to produce blocks. 

\Glspl{elected-validator} receive rewards for their participation in each \gls{batch}. The rewards are composed of a base reward and the fees of transactions included in the blocks.
To avoid competing incentives between block producers, we propose distributing the rewards proportionally to the number of \glspl{slot} a validator occupies instead of the number of blocks it produces.

Moreover, we only distribute the rewards at the end of the subsequent batch to allow for the submission of fork proofs. This ensures that we refrain from rewarding misbehaving slots.

\subsection{Punishments}\label{sec:specification:punishments}
Whether a validator is responsible for creating or continuing on a fork, creating double proposals, or casting double votes inappropriately, it is punished in a similar fashion, resulting in jail: (1) all of its \glspl{slot} are marked as punished for a period of 8 epochs, (2) its rewards will be burned\footnote{The reward is not divided among the other validators not to incentivize them to attack each other.} for the entire jail period, and (3) the validator is marked as inactive for the same period.

The exclusion from \gls{micro-block} production is effective immediately after an equivocation proof has been included in the chain. However, the validator can still participate in \gls{skip-block} and \gls{macro-block} voting to ensure the chain's progression. Marking the validator as inactive does not affect the currently \glspl{elected-validator} and only takes effect in the next validator election.

Validators can optionally submit proofs of malicious actions, such as forks or double voting. They can submit an equivocation proof from when they witness an offense until the end of the next epoch. This reporting window, spanning almost two epochs, allows validators to address and submit proofs of misbehavior. It coincides with the period validators must remain inactive to withdraw funds.

Furthermore, when a validator delays a block, (1) the misbehaving slot will no longer be considered for the \gls{slot-owner} selection, (2) its rewards will be burned, and (3) the corresponding validator is marked as inactive for the next validator election. Regardless, the slot in question is required to participate in \gls{skip-block} and \gls{macro-block} voting to avoid the protocol getting stuck. Note that we punish single slots instead of all slots of the same validator. The reasoning is that punishing multiple slots simultaneously incentivizes artificially splitting the stake into multiple validators.

\subsection{Chain Selection}\label{sec:chain-selection}
The chain selection algorithm must take into account malicious forks and \glspl{skip-block}. We use the following cumulative conditions, from highest to lowest priority, to order two chains:

\begin{enumerate}
	\item The chain with the most \glspl{macro-block}.
	\item The chain with an earlier skip block in the last \gls{batch} (in case of a tie, we check the next skip block).
	\item The chain with the most blocks.
\end{enumerate}

Still, two chains can tie on all three conditions. In that case, both chains are considered equal, and there is no clear chain to select. Thus, the next \gls{slot-owner} can build on top of either one.

\section{Synchronization}
\label{sec:syncing}
New nodes who want to join the network need a way of synchronizing with the blockchain since the only information they start with is the \textit{genesis} block, which is hardcoded into the client software.
We propose three different types of nodes that vary in the amount of data stored and needed to synchronize.
More details about the proposed synchronization methods can be found in Nimiq's state sync paper~\cite{berrang2023fast}.

\subsection{History Nodes}
History nodes need to download and verify all blocks. They are the safest option to synchronize but also the slowest, requiring the node to download vast amounts of data. Another disadvantage is that the amount of data to download grows linearly with time. Thus, the synchronization time for history nodes increases over the life of the blockchain. We expect history nodes to be run only by businesses and other public services (for example, block explorers). We leave the creation of incentives to run history nodes to the individual implementations of the protocol.

\subsection{Full Nodes}\label{sec:full-nodes}
Full nodes allow faster syncing by not downloading individual transactions but instead downloading the resulting state. On a high level, full nodes first sync to the latest \gls{macro-block} while ignoring any \glspl{micro-block} on the way. Only then do they start downloading the state and the micro blocks of the last \gls{batch}. To be able to verify the state, the blocks should contain a commitment to this state, in addition to what was outlined in \autoref{sec:block-format}. In this paper, we view the specification of the state orthogonal to the consensus protocol and leave it up to the implementation.

To sync to the latest macro block, full nodes can follow the chain of election blocks. Each election block contains a reference to its predecessor election block and is signed by the validators elected in its predecessor. Once the node reaches the last election block, it can extend its chain by the latest checkpoint block. The checkpoint block, similarly, is also referencing its predecessor election block and is signed by the respective validators.

The synchronization time for a full node still grows linearly over time but much more slowly, only adding one election block each \gls{epoch}. Furthermore, in order to save disk space, full nodes can safely prune the micro blocks of old epochs.

Note that full nodes in Albatross do not verify blocks before the current epoch apart from verifying the chain of signatures of the validators. It thus is secure as long as Tendermint's security assumption holds. 

\subsubsection{Zero-Knowledge Proofs}
Since the validation of the chain of election blocks mainly consists of simple checks and signature verifications, it can easily be incorporated into a zero-knowledge proof. With the recent advances in zero-knowledge proofs~\cite{bowe2019halo,sasson2014recursive}, the most obvious choice is to construct the proofs recursively. Each proof verifies that the current election block points back to its predecessor and is signed by two-thirds of its validator set. Additionally, it verifies the previous proof and thus concludes the correctness of the chain of election blocks. Depending on the zero-knowledge proof system employed, this can lead to a constant or logarithmic-sized proof and constant or logarithmic verification complexity.

\subsection{Light Nodes}
Light nodes are intended to be run only by end-users of the blockchain, especially in highly constrained environments like browsers and smartphones. 

In principle, light nodes can be seen as full nodes that do not download or maintain the state of the blockchain. They use the same mechanism as full nodes to sync to the latest checkpoint block. From then on, they only need to follow \gls{micro-block} headers.
This is all they need to query the blockchain state securely. Light nodes can ask full or history nodes for specific parts of the state together with a corresponding proof. These proofs can then be verified against the commitment to the state that is included in the block header (cf.~\autoref{sec:full-nodes}).

\section{Security analysis}
\label{sec:security}
We will now analyze the security of Albatross. First, we will introduce the adversarial model, and then we will discuss static and adaptive adversaries, network partitions, probabilistic finality, and transaction censorship.

\subsection{Adversarial Model}
For our security analysis, we distinguish between different types of economic actors:

\begin{itemize}
	\item \textbf{Altruistic actors:} Follow the protocol even if it is prejudicial to them.
	\item \textbf{Honest actors:} Follow the protocol as long as it is not prejudicial to them.
	\item \textbf{Rational adversaries:} Deviate from the protocol if it is profitable to them.
	\item \textbf{Malicious adversaries:} Deviate from the protocol even if it is prejudicial to them. This also includes faulty nodes. Malicious adversaries are always assumed to collude.
\end{itemize}

\subsubsection{Security Assumptions}
Based on the definitions of economic actors, we can state the central security assumption for Albatross. This assumption is derived from the fact that any reliable, deterministic Byzantine fault-tolerant algorithm is only resistant to up to a third of faulty participants~\cite{lamport1982byzantine}.

\begin{assumption}\label{assumption:slots}
Given $n=3f+1$ \glspl{elected-validator}, less or equal to $f$ of these are controlled by \textbf{colluding} adversaries.
\end{assumption}

This also means that all elected validators can be rational adversaries as long as they are not colluding and no single adversary holds more than $f$ \glspl{slot}. That is because non-colluding rational adversaries will not be able to derive any profit from the actions of others. In fact, it might harm their profits (e.g., if they do not receive rewards due to a chain stall).

In Albatross, the elected validators are chosen randomly from the larger set of active validators. We thus need to describe our assumption in terms of the percentage of the total stake controlled by an individual.

For a validator list of size $n$, if someone controls a fraction $p$ of the entire stake then the probability of him gaining control of at least $x$ validators is given by the cumulative binomial distribution:

\begin{equation}
    P(X \geq x)= \sum_{k=x}^{n} \binom{n}{k} p^k (1-p)^{n-k}
\end{equation}

For a given $n$, we now want to find the maximum $p$ such that the probability of getting assigned more than $f$ slots is negligible.

\begin{eqnarray}
    \max_p P(X > f) \leq \epsilon \\
    \Leftrightarrow \max_p P(X \geq f+1) \leq \epsilon
\end{eqnarray}

with $\epsilon > 0$.

\begin{figure}[t]
    \centering
    \includegraphics[width=.8\linewidth]{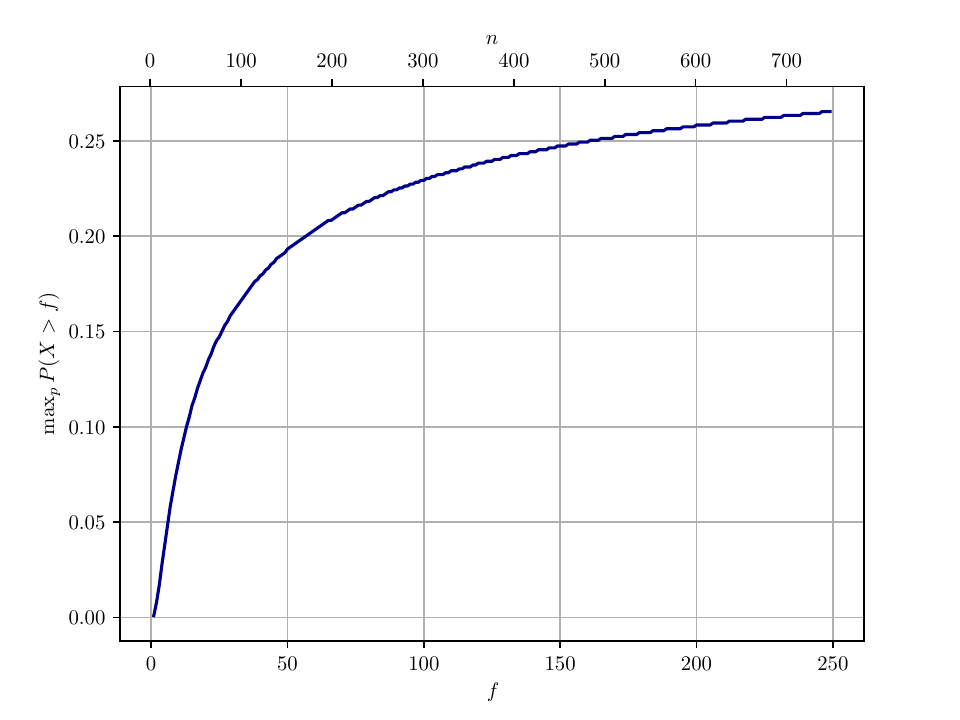}
    \caption{Maximum fraction of adversarial stake for different $n$}
    \label{fig:max_p}
\end{figure}

Setting $\epsilon=2\cdot 10^{-5}$, \autoref{fig:max_p} depicts the maximum $p$ for different values of $n$.
We can see that for a sufficiently large number of slots $n \geq 500$, the maximum fraction of adversarial stake is less or equal to $\frac{1}{4}$.

\begin{assumption}\label{assumption:stake}
Given $n \geq 500$ elected validators, less than $\frac{1}{4}$ of the total stake is controlled by colluding adversaries.
\end{assumption}

\autoref{assumption:stake} implies \autoref{assumption:slots} with overwhelming probability.
If necessary, tighter bounds for the fraction of adversarial stake can be derived for different $n$.

\subsection{Static Adversary}
First, we analyze the security of our protocol in the face of static adversaries. These adversaries can only compromise specific nodes at the beginning of the protocol. They cannot change which nodes are corrupted later on.

In our protocol, we expect the following properties to hold:

\begin{property}[Finality]\label{prop:finality}
No adversary can achieve finality on two \glspl{macro-block} at the same height.
\end{property}

\begin{property}[Probabilistic Finality]\label{prop:prob-finality}
The probability of an adversary creating a fork of length $d$ is less than $3^{-d}$.
\end{property}

\begin{property}[Liveness]\label{prop:liveness}
No adversary can prevent the chain from producing regular \glspl{micro-block} indefinitely.
\end{property}

\begin{property}[Correctness]\label{prop:validity}
No adversary can include invalid blocks in the blockchain.
\end{property}

We bundle these properties into our security definition and show that our protocol fulfills these given our security assumption.

\begin{theorem}[Protocol Security]\label{theorem:security}
Given that \autoref{assumption:slots} holds at any point in time, Albatross satisfies the properties \autoref{prop:finality}--\autoref{prop:validity} for static adversaries.
\end{theorem}

In the following, we will argue the validity of \autoref{theorem:security} for each property individually before we explain what happens if our assumption fails.

\subsubsection{(Probabilistic) Finality}\label{sec:finality}
When receiving a transaction, knowing whether the transaction can be rever because of a fork is essential.

Macro blocks are being produced with the Tendermint protocol. Thus, \autoref{prop:finality} directly follows from the properties of Tendermint.

In contrast, micro blocks cannot guarantee the same level of finality as they are only signed by a single validator. However, Albatross offers strong probabilistic finality.

\begin{proof}[Proof of \autoref{prop:prob-finality}]
Finality can only be broken through a blockchain reorganization. According to the chain selection rules in \autoref{sec:chain-selection}, this can only happen by (1) producing a fork that is long enough or (2) by producing a \gls{skip-block} at an earlier block height.

Neither of these options is in the interest of an altruistic or honest actor, as they both incur a punishment. Rational and malicious adversaries are only allowed to hold up to $f$ \glspl{slot}, and hence they are unable to produce a skip block at an earlier height. They can neither produce a skip block by themselves (requires $2f+1$ signatures) nor can they withhold one (requires $f+1$ signatures).

Let $F$ denote the length of a fork.
To produce a fork of length $F=d$, the adversaries need to be \glspl{slot-owner} for all $d$ blocks since altruistic or honest actors would resolve the fork.
Assuming the worst-case scenario of the adversaries holding $f$ slots, we thus get:

\begin{equation}
\begin{split}
	P(F = d) &=\left( \frac{f}{n} \right)^d \\
 &= \left( \frac{n-1}{3n} \right)^d = \left( \frac{1}{3} - \frac{1}{3n} \right)^d\\
 &< \left( \frac{1}{3} \right)^d=3^{-d}
\end{split}
\end{equation}
\end{proof}

This means that the probability of a transaction being reverted due to a fork decreases exponentially. A client can easily calculate the minimum probability that a transaction is final by considering the number of blocks built on top of the block that includes the transaction.

\autoref{table:finality} demonstrates that a certainty of 99.9\% is reached after only six blocks (including the block containing the transaction).

\begin{table}[h]
	\label{table:finality}
	\caption{Probability of a transaction being final after \emph{n} blocks}
	\centering
	\begin{tabular}[c]{l||c|c|c|c|c|c}
		Blocks & 1 & 2 & 3 & 4 & 5 & 6 \\
		\hline
		  Prob.~of finality & 0.666 & 0.889 & 0.963 & 0.988 & 0.996 & 0.999 \\
	\end{tabular}
\end{table}

\subsubsection{Liveness}
A static adversary should not be able to stop the blockchain from progressing. While we express \autoref{prop:liveness} in terms of micro blocks, this property also affects macro blocks: if an adversary is able to prevent macro blocks from being produced, it also inhibits the production of micro blocks. We thus consider both cases in our analysis.

For macro blocks, we again rely on the properties of Tendermint. An adversary can neither prevent a macro block from being proposed (any attempt will lead to a new round and new proposer) nor can it prevent $2f+1$ votes from forming (given \autoref{assumption:slots}).

In the case of micro blocks, a sufficiently large delay entails the production of a skip block (see \autoref{sec:specification:skip-blocks}). Due to \autoref{assumption:slots}, the adversary does not have enough voting power to prevent the skip block from being added to the blockchain. The next block will have a newly elected slot owner. Since the worst-case probability of an adversary being the slot owner in $d$ consecutive blocks is $3^{-d}$ (cf.~\autoref{sec:finality}), the chances of indefinitely preventing the production of a regular micro block are negligible. Moreover, each skip block results in the adversary losing one more micro block production slot.

Thus, we can mitigate any attempt to stop chain progress and will eventually produce a \gls{micro-block}.

\subsubsection{Correctness}
In Albatross, altruistic or honest actors will always ignore invalid blocks. Thus, we must show that no rational or malicious adversary can create an accepted chain with such blocks.

In order to finalize a chain with invalid blocks or finalize an invalid macro block, Tendermint requires $2f+1$ votes. However, from \autoref{assumption:slots}, it follows that such a macro block can never be approved by the adversary alone. Altruistic and honest actors will ignore the proposal and vote for a new round. Rational adversaries with less than $f$ slots will behave the same, knowing they cannot finalize the block by themselves and thus do not profit from it. Hence, an adversary cannot finalize such a chain.

In optimistic mode, when an invalid micro block is produced, the honest and altruistic actors will discard it and, after the timeout, create a skip block. \autoref{assumption:slots} ensures that the rational and malicious adversaries do not have enough voting power to prevent the skip block from being added to the blockchain. 

Similarly to the macro blocks, an adversary cannot create and include invalid skip blocks because it lacks voting power.

\subsection{Adaptive Adversary}
Next, we discuss adaptive adversaries. While an adaptive adversary can change which nodes are compromised at any given point in time, it is still restricted in the total number of nodes that can be simultaneously compromised.

We only need to consider cases where the adversary can corrupt at most $f$ validators. An adversary that can corrupt more than that can already compromise the consensus algorithm in the static case.

For an adaptive adversary to break \autoref{prop:finality} and \autoref{prop:validity}, several options exist. Such an adversary, for example, could compromise nodes round-robin and produce the required signatures. Another attack that might be possible for an adaptive adversary is a \emph{long-range} attack, in which the adversary compromises old keys of validators (potentially even bribing them) to finalize an alternate \gls{macro-block} in the past. To any implementation of the Albatross protocol, we propose to counter such attacks through a combination of operational security measures and off-chain defenses.

To break \autoref{prop:prob-finality} or \autoref{prop:liveness}, the simplest way is for the adaptive adversary to always corrupt the current \gls{slot-owner}. To achieve this, however, the adversary needs to know the next slot owner prior to the production of this block. Since the slot owner is only selected in the previous block, and it does not require any interaction with other validators to produce a block, \emph{the adversary must learn who will be the slot owner before the slot owner itself}.

Naively, the attacker can create many nodes in the network so that he can receive a block before the next slot owner does. That will give him an antecedence of roughly the \emph{block propagation time} over the slot owner. The attacker would thus need to compromise the next slot owner during this short period of time.

Another strategy for the adversary is to wait for its turn to produce a block. Now the adversary can learn the identity of the next slot owner before publishing the block. Using this technique, the attacker can have an antecedence equal to the \gls{skip-block} timeout, during which the attacker needs to compromise the next slot owner.

Although strictly speaking, an adaptive adversary can corrupt nodes instantly, a more realistic model would consider the time to corrupt a node. Independently of the strategy used, in Albatross, an adaptive attacker would need to corrupt nodes on the order of seconds. Since Albatross is a decentralized system with heterogeneous nodes, systematically corrupting them would be not only complex but also costly, making such an attack unlikely.

\subsection{Network-level Adversaries}
So far, we have only discussed adversaries that are participating in the Albatross protocol. However, it is equally important to consider faults or malicious actions on the network level, specifically those causing network partitions.

\begin{property}[Resistance to Network Partitions]
In the event of a network partition, Albatross guarantees consistency of \glspl{macro-block} and prevents arbitrarily long forks.
\end{property}

From the \emph{CAP theorem}~\cite{gilbert2002brewer}, we know that when suffering a network partition, a distributed system can only maintain either consistency or availability. Tendermint favors consistency over availability and will stop in the presence of a network partition. Albatross also favors consistency but can potentially still produce a few \glspl{micro-block} before stopping.

Note that if the network is split into two parts, it is possible for one of them to contain the \glspl{slot-owner} of the following $z$ blocks. In this case, these blocks will be produced before a \gls{skip-block} is attempted at $z+1$. However, skip blocks can only ever be produced when $2f+1$ \glspl{slot} are in the same partition.

As soon as the network partition ends, Albatross will quickly resume its normal operation. The $z$ blocks produced by one part will be accepted by the remainder, and then the nodes can start producing blocks from there.

It is worth noting that if one of the parts has $2f+1$ or more \emph{rational}, honest, and altruistic validators, then Albatross is potentially able to continue normally, preserving both consistency and availability.

\subsection{Attack Scenarios when Assumptions Fail}
Finally, we will give some intuition on the possible attacks should our assumptions not hold. In this case, the main difference between pure Tendermint and Albatross is that, in Tendermint, all blocks are \emph{final}, so all transactions are irreversible as soon as they get published in a block. In contrast, transactions in Albatross have only \emph{probabilistic finality} within an \gls{epoch}, although the probability of reversibility is exponentially decreasing.

Depending on the number $x$ of validators controlled by the adversary, there are two relevant cases outside of our assumptions:

\begin{itemize}
	\item $f < x \leq 2f$: The adversary can delay the network indefinitely by stalling the \gls{skip-block} protocol. Also, the adversary can revert arbitrarily long chains within an epoch by exploiting the skip block protocol and the chain selection rules.

    Let us imagine that the adversary controls $f+1$ validators. As soon as the adversary is the \gls{slot-owner} for a \gls{micro-block}, it produces that block but withholds it until receiving skip block messages from at least $f$ other \glspl{slot}. Then, the adversary will release its block. By combining the received skip block messages with its own, the adversary can make the chain revert to the skip block as long as no \gls{macro-block} has yet been produced.
	\item $x > 2f$: The attacker has complete control over the network and can delay it, create forks and publish invalid blocks.
\end{itemize}

\section{Protocol Upgrades}
\label{sec:upgrades}
In traditional blockchain systems, upgrading the protocol poses the problem of permanent chain forks.
Albatross prioritizes \emph{consistency} over \emph{availability}. Thus it is not possible to have chain splits during \glspl{macro-block}~\cite{gilbert2002brewer}. We propose to only upgrade during election blocks for simplicity. This also generalizes our upcoming analysis and makes it applicable to other Tendermint-based and PBFT-like consensus algorithms.

A naive upgrade mechanism would be to propose an election block with the updated protocol and to leave the decision to the Tendermint voting. This approach comes with a problem: non-upgraded validators remain in the system and could be elected for the next \gls{epoch}, potentially leading to a stall. However, simply deactivating those validators can lead to an increased proportion of malicious validators.

To address these issues, we propose the protocol in \autoref{alg:upgrade}.

\begin{algorithm}
\caption{Upgrade Mechanism}\label{alg:upgrade}
\begin{algorithmic}[1]
\State $v_\textsf{old}, v_\textsf{new} \gets$ versions pre- and post-upgrade
\State $\mathcal{V}_{v_\textsf{old}} \gets$ validators that did not signal $v_\textsf{new}$
\State $p \gets$ proportion of stake supporting upgrade
\State $t \gets$ supporting threshold
\If{$p > t$}
    \State $b \gets$ \Call{ProposeBlock}{$v_\textsf{new}$}
    \If{$b$ is finalized}
        \State \Call{DeactivateValidators}{$\mathcal{V}_{v_\textsf{old}}$}
    \Else
        \State \Call{ProposeBlock}{$v_\textsf{old}$}
    \EndIf
\Else
    \State \Call{ProposeBlock}{$v_\textsf{old}$}
\EndIf
\end{algorithmic}
\end{algorithm}

This new protocol requires validators to signal their intention to support the upgrade on-chain. Besides stopping the blockchain from halting by deactivating non-upgraded validators, it also limits the proportion of malicious validators after an adopted upgrade through a \emph{supporting threshold} $t$.

Since we require the proportion of malicious validators' stake to be smaller than $\frac{1}{4}$ after the upgrade (see \autoref{assumption:stake}), we can now determine the maximum fraction of malicious validators we can have before the upgrade.

\begin{theorem}[Upgrade Security]
Let $p$ be the exact proportion of nodes supporting an upgrade.
For the protocol in \autoref{alg:upgrade} to adhere to our security assumption \autoref{assumption:stake} post-upgrade, the proportion of malicious stake before the upgrade must be less than $\frac{1}{4}\cdot p$.
\end{theorem}

\begin{proof}
Let $total_\textsf{pre}$ and $total_\textsf{post}$ be the total stake before and after the upgrade.
Let $mal_\textsf{pre}$, $mal_\textsf{post}$ be the absolute malicious stake before and after the upgrade. 
$p$ is the proportion of the pre-upgrade stake that applied the upgrade, thus $total_\textsf{post}=p \cdot total_\textsf{pre}$.

To maximize the amount of malicious stake after the upgrade, we assume that all malicious validators supported the upgrade, hence $mal_\textsf{pre} = mal_\textsf{post}$.

Now, if we require the proportion of malicious stake after the upgrade $\frac{mal_\textsf{post}}{total_\textsf{post}} < \frac{1}{4}$, it follows that:
\begin{align*}
    \frac{mal_\textsf{post}}{total_\textsf{post}} &< \frac{1}{4}\\
    \Leftrightarrow \frac{mal_\textsf{pre}}{p\cdot total_\textsf{pre}} &< \frac{1}{4}\\
    \Leftrightarrow \frac{mal_\textsf{pre}}{total_\textsf{pre}}
    &< \frac{1}{4} \cdot p\\
\end{align*}

Hence, before the upgrade, the proportion of malicious stake must not be more than $\frac{1}{4} \cdot p$.
\end{proof}

Our analysis is valid as long as the proportion of upgrading validators is greater or equal to our threshold $p\geq t$.

For Albatross specifically, the stake-based threshold $t$ must also be larger than $\frac{1}{2}$. Otherwise, the nodes that oppose the upgrade likely hold more than $2f+1$ \glspl{slot} and can refuse to finalize any block containing the upgrade with non-negligible probability\footnote{This bound can be derived similarly to \autoref{assumption:stake}.}. However, if the threshold $t$ is much larger than $\frac{1}{2}$, malicious validators can stop an upgrade by simply refusing their support. In the limit, when $t=1$, a single malicious validator can force the chain never to be upgraded.

Our security argument for upgrades is generic to the extent that an analogous analysis can be carried out on other PoS chains.

\begin{figure*}[t]
    \centering
    \includegraphics[width=.7\linewidth]{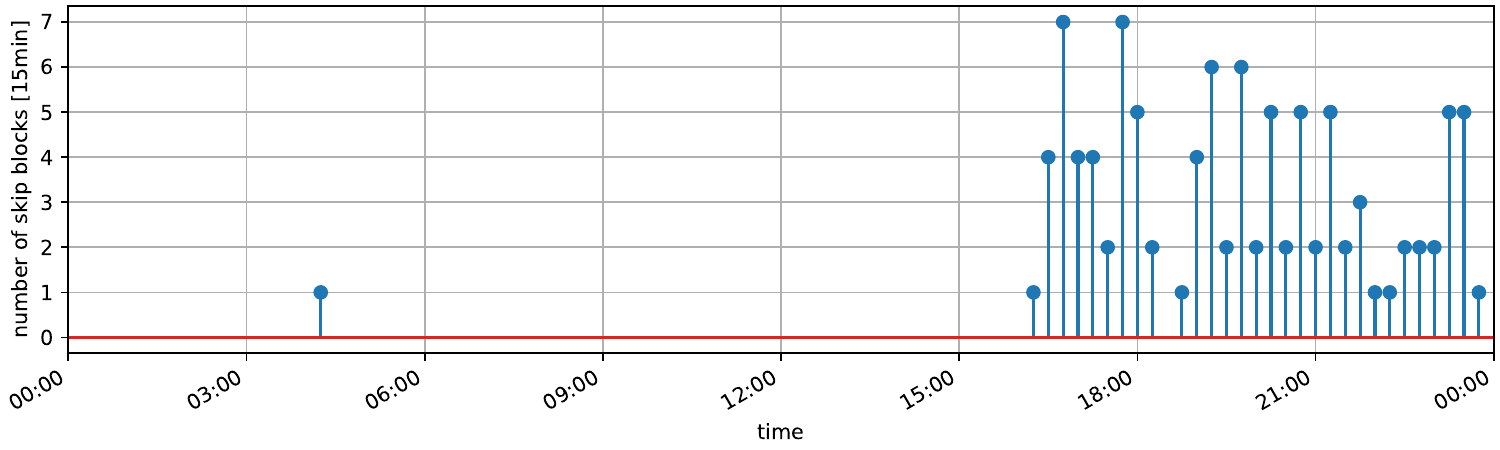}
    \caption{Distribution of skip blocks in real-world measurements}
    \label{fig:skip-blocks}
\end{figure*}

\section{Theoretical Protocol Performance}
\label{sec:theoretical-performance}
In this section, we give a brief theoretical analysis of the performance of Albatross. We show that, in the \emph{optimistic} case, it achieves the theoretical limit for single-chain PoS algorithms, while in the \emph{pessimistic} case, it still achieves satisfactory performance.

\subsection{Optimistic case}
The best case is when the network is synchronous (all messages are delivered within a maximum delay $d$), the network delay $d$ is smaller than the timeout parameter $\Delta$, and all validators are honest.

\Glspl{macro-block} have a message complexity of $\mathcal{O}(n^2)$, since they are produced with Tendermint. However, they constitute only a small percentage of all blocks, so the overall performance is mainly correlated with the \gls{micro-block} production.
Approaches such as Handel~\cite{bgassat2019handel} can be used to reduce the macro block message complexity.

Micro blocks have a message complexity of $\mathcal{O}(1)$. In fact, they only require the propagation of the block. If we ignore the time spent verifying blocks and transactions, the latency is equal to the block propagation time, which is on the order of the network delay $d$.

In conclusion, Albatross, in the optimistic case, produces blocks as fast as the network allows it.

\subsection{Pessimistic case}
If we relax some of the assumptions made for the optimistic case, Albatross still has a performance superior to Tendermint. There are two different cases that we will consider:

\subsubsection{Malicious Validators}
The worst case, while still maintaining security, is a scenario with $f$ validators being malicious and refusing to produce blocks. In this case, we can expect one \gls{skip-block} every three blocks. The skip block protocol requires $\mathcal{O}(n)$ messages and waiting for a timeout. So, in this case, the message complexity will be $\mathcal{O}(n)$, and the latency will be on the order of $\Delta$.


\subsubsection{Partially Synchronous Network}
Under partial synchrony, the network oscillates between periods of network asynchrony and synchrony. In this case, it is possible to completely halt the progress of the blockchain while the network is asynchronous. However, Albatross will return to normal operations when the network becomes synchronous again.

\section{Implementation \& Performance Evaluation}
\label{sec:impl}
We provide an open-source Rust implementation\footnote{\url{https://github.com/nimiq/core-rs-albatross}} of the Albatross protocol as part of the Nimiq PoS blockchain system. The implementation can also be compiled to WebAssembly and runs in web browsers. In the following section, we describe the instantiation of our implementation and further additions that go beyond Albatross' specifications.

Our implementation uses Schnorr signatures~\cite{schnorr1990efficient,schnorr1991efficient} to sign transactions. Thus every address is associated with a Schnorr public key. When signing up in the validator registry, validators provide an address, a second independent Schnorr public key, and a BLS~\cite{boneh2004short} public key. The second Schnorr key is used for signing micro blocks and random seed production (using VXEdDSA~\cite{perrin2016xeddsa}). Having two independent Schnorr keys is done for operational security, while the BLS key enables efficient signature aggregation for skip block and macro block voting. We also leverage Handel~\cite{bgassat2019handel} to reduce the message complexity during aggregation.

Our \glspl{epoch} consist of 720 \glspl{batch} with 60 blocks each. The number of \glspl{slot} in each epoch is set to 512 and thus above what is required for \autoref{assumption:stake}.

The blockchain state is modeled as a Merkle-Radix trie, which facilitates the synchronization of the blockchain~\cite{de2022overview}. We offer a small set of hard-coded contracts, such as Hash-Time-locked contracts and vesting contracts. The validator and stakers registry is also implemented as a hard-coded contract.

We extend the Albatross protocol by a fixed block separation time of 1 second. This means that validators always aim for blocks to be spaced in one-second intervals, giving the protocol more predictable finality and allowing end-users in constrained environments to run light nodes (e.g., browsers and smartphones). If validators do not adhere to the block separation time, their rewards will be reduced.

\subsection{Real-World Performance}
We run our implementation in a public test network, and instructions to connect can be found on the project's GitHub page. The network is run in a primarily non-adversarial environment, and we collected data over a period of 24 hours on May 9, 2023. The Nimiq PoS test network was launched end of March 2023 and, since then, produced more than 1{,}296{,}000 blocks (60 epochs). The chain also accumulated more than 4{,}000{,}000 transactions.

\autoref{table:performance} shows statistics on the block production times and the number of connected peers measured on a small subset of nodes in the network. The block production times are closely following the implemented block separation time of 1 second and only rarely deviate from it. The number of connected peers is mostly stable but slightly fluctuates due to the public nature of the network. Overall, we can see a stable network behavior that is only capped by the block separation time and hence supports our theoretical analysis in \autoref{sec:theoretical-performance}.

We also investigate the production of \glspl{skip-block} and fork proofs during the same time period. While -- due to the non-adversarial nature -- no forks were constructed, we were able to observe the production of skip blocks. These skip blocks mainly originated because two validators went offline during that time. \autoref{fig:skip-blocks} depicts the distribution of skip blocks over the measurement period. Each bar represents the number of skip blocks produced in a 15-minute interval. The two validators that went offline at around 16:00 occupied multiple slots, so it took several skip blocks to punish all of their slots (cf.~\autoref{sec:specification:punishments}). During the whole period, the network continued to operate normally.

\begin{table}[t]
	\label{table:performance}
	\caption{Nimiq PoS performance measurements}
	\centering
	\begin{tabular}[c]{l||c|c|c|c}
             & avg. & std. & min. & max. \\
            \hline
            \hline
		block time & 996.59 ms & 18.48 ms & 450 ms & 1020 ms \\
		\hline
		  connected peers & 18.77 & 4.82 & 12 & 25 \\
	\end{tabular}
\end{table}

\section{Related Work}
Albatross is modeled after speculative BFT algorithms such as Zyzzyva~\cite{kotla2009zyzzyva} and takes inspiration from other consensus algorithms like Byzcoin~\cite{kokoriskogias2016byzcoin} and Bitcoin-NG~\cite{eyal2015bitcoinng}.

Similarly to Albatross, Byzcoin and Bitcoin-NG distinguish between different types of blocks called key blocks and micro blocks. Byzcoin uses key blocks to replace one validator at a time, while Albatross uses election blocks to replace the complete validator set. Moreover, in Albatross, both blocks are published on the same blockchain, while Byzcoin models them as separate chains.

Furthermore, Albatross' use of VRFs to generate a chain of random seeds is inspired by Algorand's design~\cite{chen2016algorand}. Algorand, however, does not select a single block producer using the random seed. Instead, blocks are proposed by the validator that can prove to have the lowest random value output by his VRF.

\section{Conclusion}
\label{sec:conclusion}
This paper specifies and analyzes Albatross, a novel consensus algorithm inspired by speculative BFT algorithms. Our protocol provides periodic provable finality with intervals of strong probabilistic finality. Its design positions Albatross as a high-performing blockchain protocol.

Albatross distinguishes between macro blocks, committed through Tendermint, and optimistically produced micro blocks. This design also simplifies synchronization for devices with low hardware specifications. Thus, our open-source implementation even runs in browsers and smartphones.

We also prove that our protocol is secure using standard assumptions of BFT consensus algorithms and relate these assumptions to the maximum amount of malicious stake we support.

Finally, we analyze the performance of our protocol, both from a theoretical perspective and through real-world measurements based on our implementation. Our measurements support that the performance of our protocol is limited only by the network.

\printbibliography

\printglossary

\end{document}